# What is the Nature of Chinese MicroBlogging: Unveiling the Unique Features of Tencent Weibo

Daifeng Li*, Jingwei Zhang‡, Jie Tang*, Gordon Guo-zheng Sun⋄, Ying Ding♯, Zhipeng Luo↘
*Dept. of Computer Science and Technology, Tsinghua University, Beijing, China
‡Depart. of Electronic Engineering, Tsinghua University, Beijing, China
⋄Tencent Company, Beijing, China
♯School of Library and Information Science, Indiana University Bloomington, IN, USA
↘Beijing University of Aeronautics and Astronautics, Beijing, China
ldf3824@yahoo.com.cn, iceboal@gmail.com, jery.tang@gmail.com, gordon.gzsun@gmail.com
dingying@indiana.edu, patrick.luo2009@gmail.com

## ABSTRACT

China has the largest number of online users in the world and about 20% internet users are from China. This is a huge, as well as a mysterious, market for IT industry due to various reasons such as culture difference. Twitter is the largest microblogging service in the world and Tencent Weibo is one of the largest microblogging services in China. Employ the two data sets as a source in our study, we try to unveil the unique behaviors of Chinese users.

We have collected the entire Tencent Weibo from 10th, Oct, 2011 to 5th, Jan, 2012 and obtained 320 million user profiles, 5.15 billion user actions. We study Tencent Weibo from both macro and micro levels. From the macro level, Tencent users are more active on forwarding messages, but with less reciprocal relationships than Twitter users, their topic preferences are very different from Twitter users from both content and time consuming; besides, information can be diffused more efficient in Tencent Weibo. From the micro level, we mainly evaluate users' social influence from two indexes: "Forward" and "Follower", we study how users' actions will contribute to their social influences, and further identify unique features of Tencent users. According to our studies, Tencent users' actions are more personalized and diversity, and the influential users play a more important part in the whole networks.

Based on the above analysis, we design a graphical model for predicting users' forwarding behaviors. Our experimental results on the large Tencent Weibo data validate the correctness of the discoveries and the effectiveness of the proposed model. To the best of our knowledge, this work is the first quantitative study on the entire Tencentsphere and information diffusion on it.

## Categories and Subject Descriptors

H.2.8 [**Database and Management**]: Data Mining; J.4 [**Computer Applications**]: Social and Behavioral Science

## 1. INTRODUCTION

Tencent Weibo, the biggest Micro-Blogging website not only in China, but also in the world, is considered powerful tool to change people's traditional communication styles in China. The basic functions of Tencent is similar with Twitter, which include "Follow", "Create tweets", "Retweet", "Reply", "Mention", "hastag" and et al. In Tencent Weibo, we call a tweet as weibo, "Retweet" as "Forward","hastag' as "Topic". Different with Twitter, Tencent Weibo provides more functions to help online users to build their own personalized media center for both obtaining and diffusing information; the new functions include "Mails", "Comments", "figures and videos sharing" and etc. "Comments" provide specific space for users to discuss on a certain topic; "Mails" could provide mail communications between two users. All the assignments above could help users to participant in group discussions in a more easier way and could better express their opinions, users could also spread interesting information more flexible. We call all users' behaviors through those functions as users' actions, the actions of Tencent users result in a highly sensitive information system, which means that useful information will spread rapidly and cover millions of nodes level as the formal information is announced even before.

Researches of large-scaled data analyses have been widely studied and applied in many famous micro-blogs, such as Twitter and Facebook. Those researches take deep insight into topological features and dynamic trends of evolution, and their research results are an important complement to current social network theory, but they seldom consider specific social networks that are embedded in different culture, economics and political background. In this study, we focus on Tencent Weibo, the most famous of Chinese micro-blog, which has more than 320 millions users, there is no previous study of such a social network, which consist of such a huge amount of users from the same country. What these users are talking about every day, how they construct their on-line structures, and how they share and spread information is the essential information we hope to identify. The main purpose of this paper is thus to study the statisti-cal features of Tencent Weibo, which could help us better understand its unique capabilities and components. A set of experiments are designed to make a systematic research from both macro and micro levels; from the macro level, we investigate whether Tencent Weibo is a new media center or a communication network, and from the micro level, we investigate the behavioral features of Tencent users and compare it with Twitter. We apply paralleled Topic Model, People-Rank, CRFs(Conditional Random Fields) and other Map-Reduce based statistical algorithms to support our researches. Our main contributions are included as follow:

- We analyze users' actions in Tencent weibo, and find that compared with Twitter, Tencent weibo is more likely to be a personalized media center, and the features of users' actions are very different with that of Twitter.

- The hot topics talked in Tencent weibo are very different with that of Twitter.

- Compared with twitter, Tencent has a more complex network and more active users.

- Similar with Twitter, the number of "Forwards" and "Followers" is related with users' actions in Tencent, and the relationships are only effective under a certain extent.

- To verify the practical value of our discovers in Tencent Weibo, we propose a predictive model to illustrate the applications of forward analysis, the results show that statistical analysis of Tencent weibo could be effectively applied to direct the design of weibo services.

This paper is organized as follows: Section 2 introduces the related work; Section 3 describes the data collection; Section 4 makes an analysis from the macro level; Section 5 makes an analysis from the micro level; Section 6 propose an application research; and Section 7 is the conclusion.

## 2. RELATED WORK

Micro-Blogs, as a new style of online social network and social media, has recently attracted more and more attention. For example, Twitter, as one of the world's biggest micro-blogs websites, has been widely studied. Users' behaviors and social relations are always the hot research topics, Perra et. al took Twitter as experiment objects, constructed an activity-driven model to describe the structure features of the highly dynamic network [17]; Jie et. al studied on how different social ties will influence the propagation of information in Twitter [20]; Java et. al investigated how users communicate with each other and generate communities in Twitter [11]; Krishnamurthy, et. al used follower-followee relationships to study users' characteristics [13]. The influence of Twitter users is another attractive topics, Wu et. al defined and applied new features to observe the behaviors of different type of influential users in Twitter [?]; Cha et. al measure influence in Twitter from three aspects: number of "Followers", "Forwards" and "Mentions" [6]; Some researchers focused on using the analysis results for real applications, Hopcroft et. al studied the features of "Reciprocal" in Twitter and made prediction on users' behaviors [9]; Peng et. al studied the features of "Retweet" in Twitter and applied CRF to make "tweets" recommendations [16]; Zhao et. al studied the potential reasons for users' behaviors in Twitter and make predictions [21]; Other interesting researches are mainly focused on what users are talking in Twitter, for example, HP Labs made primary comparison between Twitter and Sina Weibo to analyze their daily topics; Banerjee et. al detected topic interests of Twitter users by analyzing their posted tweets [1] [2]; Other researches revealed Twitter's social sensor and prediction, such as [3] [10] [19] and [18]. Several researches have also systematically studied Twitter, Kwak et. al applied statistical methods to analyze its features from both social network and social media angles [14]. Similar to Kwak's work, our research also marks the first look at the entire micro-blogs components of Tencent, and makes comparison with Twitter. This is meaningful for current researches of social networks, because few studies have been done to study specific micro-blogs with different cultural, economics and political backgrounds. The researches of Tencent Weibo's more than 320 million Chinese users may provide important information that complement existing studies. A systematic attempt at characterizing unique features of Tencent Weibo is designed, the experiment takes deep insight into users' online actions, information propagation and dynamic content analysis by making comparison analysis with Twitter.

## 3. DATA COLLECTION

We collect the entire data set of Tencent Weibo from 7 Oct 2011 to 5 Jan 2012 day by day, wherein each day contains 40 million 150 million users' actions; we use a servers cluster with 36 machines to store all data sets in HDFS system, wherein each machine contains 15 Intel(R) Xeon(R) processors (2.13GHZ) and 60G memory. In order to obtain a high quality experiment data set, we also make rules to delete spam users from Tencent Weibo. The summarization of the final experiment data set is listed in Table 1.

The data collection is mainly based on users' all actions and their social relations in Tencent Weibo. We build profiles for each user as $UserX$={$Uname, Followers, Followees, Weibos$}, where $Uname$ is the ID of each user, $Followers$ is the collection of each user's followers, $Followees$ is the collection of each user's followees, $Weibos$={$id, type, content, parent, root, time, location$} records all weibos, which current user has created. We assign a unique $id$ for each weibo; $type$ includes "Original Create", "Forward", "Reply", "Comment", "Mail", "Mention"; $content$ contains texts, figures and videos; $time$ and $location$ are to record when and where current weibo was generated; assume we have two weibos $A$ and $B$, if $B$ is a "Reply" or "Comment" or "Forward" of $A$, then we call $A$ is $B$'s $parent$; if another weibo $C$ is $parent$ of $A$ and $C$ does not have $parent$, then $C$ is root of $B$. Besides, we also collected a small data of all actions and relationships of 110 thousands Twitter users from 12 OCT to 23 DEC by using Twitter API, the data set includes 293,386 following relationships, 9,376,500 original created tweets and 13,277,043 retweets. The data set is used as a aid for some comparison between Tencent and Twitter.

## 4. MACRO LEVEL ANALYSIS

### 4.1 Personalized Media or Social Circles

In this section, we investigate users' main preferences in Tencent Weibo. These preferences can be described as whether users like to build a personalized media center or a social circle in Tencent Weibo. The evaluations are mainly based on users' actions, "Original Create" and "Forward" can be considered as creating and sharing information from a personalized media center, while "Reply", "Mention", "Mail" and "Comment" can be considered as social communications. "Reciprocity" is also an important index to measure the willingness of users to generate social circles. Thus we evaluate all users' actions mainly from three aspects: Power Law Growth, Exponent Distribution and Reciprocity.

Table 1: The Summarization of Tencent Micro-Blogging from 2011 Oct to 2012 Jan

| Items | Users | Original | Forwards | Replies | Comments | Mails | Mention |
|---|---|---|---|---|---|---|---|
| Total Number | 326,497,021 | 3,607,924,594 | 1,026,243,542 | 43,658,122 | 299,354,146 | 174,440,376 | 2,347,927 |

### 4.1.1 Power Law Growth of Users' Actions

Growth analysis is applied to observe the growth rate of users' actions; the analysis could help to better understand users' actions in Tencent Weibo. Growth analysis are widely used in many websites such as Del.icio.us, Flickr and Youtube [15]. We design similar experiments to evaluate Tencent Weibo, the results are exhibited in Figure 1.

In Figure 1, three sub-figures illustrate the increased features of three main users' actions: "Original Create", "Forward" and "Reply") along the intrinsic time $dt$. Intrinsic time $dt$ is defined as the increased number of total users' actions during a fixed time interval. In this experiment, we assign the time interval as an hour, all the increase actions along the intrinsic time closely follow a power-law (straight line in a log-log plot) across the entire time line. The dash black lines are provided as an aid for observation, with increase exponent $gamma$ as 0.9826, 0.9189 and 0.7848 respectively, which means that users in Tencent Weibo prefer to express their opinions by creating new weibos than by communicating with others. According to the statistical data from sysomos company [1], the exponent of "Original Create", "Forward" and "Reply" in Twitter is around 0.9224, 0.8387, 0.9060 respectively. The result shows that Tencent users have higher activities to create and share new messages than Twitter, while Twitter users tend to reply rather than forward messages. While compared with other social networks, such as Del.icio.us [5], Flickr and Youtube [15], the exponent of which is around 0.8. Users tend to propagate existing weibos firstly by using "Forward" actions, while "Reply" actions gain the lowest scores. We also calculate the exponent of "Comment", "Mention" and "Mail", the results are bigger than "Reply" and smaller than "Forward". About 28% of original created weibos are related with daily sentiment. About 70% of original created weibos concerned to a certain topic, such as entertainment, economics, politics, science and etc, among which entertainment gains the highest popularity. About 64% of original created weibos contain pictures or videos; different with Tencent, Twitter users prefer events with hot economic, politic and science topics. The small Figures embedded in each Figure show the actions increasing along the real time, all three actions exhibit linear increases. According to this analysis, we can summarize that Tencent users prefer to create and share information rather than to communicate with others, while the communication actions take up a total of around 30 % of all the actions.

### 4.1.2 Exponent Distribution of Users' Actions

To further confirm our conclusions, we design the following experiment to investigate how users from different active levels organize their actions. We select the top 100,000 active users, 75% 100,000 ranked normal users based on their total number of actions, and then calculate their exponent distributions for different type of actions. The formula for calculating activity exponent is as follow:

$$Exponent_{type}(user_i) = \frac{log(F_{max}^{type})}{log(F_{max}^{all})} \quad (1)$$

In Equation 1, $type$ means different action types; $F_{max}^{type}$ means the final number of current type of actions of user $user_i$. By applying that equation, we draw the Exponent distributions of all actions (Original, Reply, Forward and Mail) of all selected users.

In Figure 2, the left sub figure can be seen as the exponent distributions of active users and the right one is the exponent distributions of normal users(75% rank). For the active users on the left sub figure, around 40 % to 50 % users have a high preference for one main action (such as users only create "Original" weibos, or users only forward others), and the remaining users prefer to assign a different proportion for their different actions, while most of them prefer "Original" over others (the exponent value between 0.6 and 0.8). We could also observe that "Original" and "Forward" have an increase trend; "Mail", "Reply", "Mention" and "Comment" have a high consistency of decrease trends. For normal users on the right sub figure, similar results could also be observed, it seems that normal users have higher activities for creating new weibos, lower activities for social communication. This phenomenon illustrates that Tencent users tend to build a personalized media center rather than create social circles; while highly active users seem to have higher activities to communicate with others, and normal users are more likely to create and share information with others.

### 4.1.3 Reciprocity Analysis

In this section, we analyze how Tencent users generate their Reciprocity relationships during the three months of data collection. We deploy the $X$ axis as the number of "Reciprocity Friends", while the $Y$ axis serves as the number of users, who generate the same number of "Reciprocity Friends" during those three months, and draw the reciprocity distributions in Figure 3.

As can be seen in Figure 3, the Reciprocity satisfies power-law distributions, with the exponent as -0.5261. The average Reciprocity number is 2.7342, which means that a user may generate an average of 2.7342 reciprocity relationships with other users. The total percentage for the three months is about 0.7% compared with the total number of "Follows" actions, which shows that after almost two years of development(Tencent Weibo was formally online on 1st April 2010), the increase of Reciprocity is quite small, compared with the increase of "Following" relationship in Figure 10, where the average following actions is 64. Based on this observation, we find users are more likely to follow new users who they are more interested in, but not their followers. One main reason for this is that Tencent Weibo is based on user groups of the Tencent QQ(the biggest online chatting services in China), which will bring social relationship of QQ users to Tencent Weibo once users accept services from Weibo; in another aspect, users could communicate with each other by adopting other more efficient Tencent tools such as Tencent QQ.

---
[1] http://www.sysomos.com/insidetwitter/

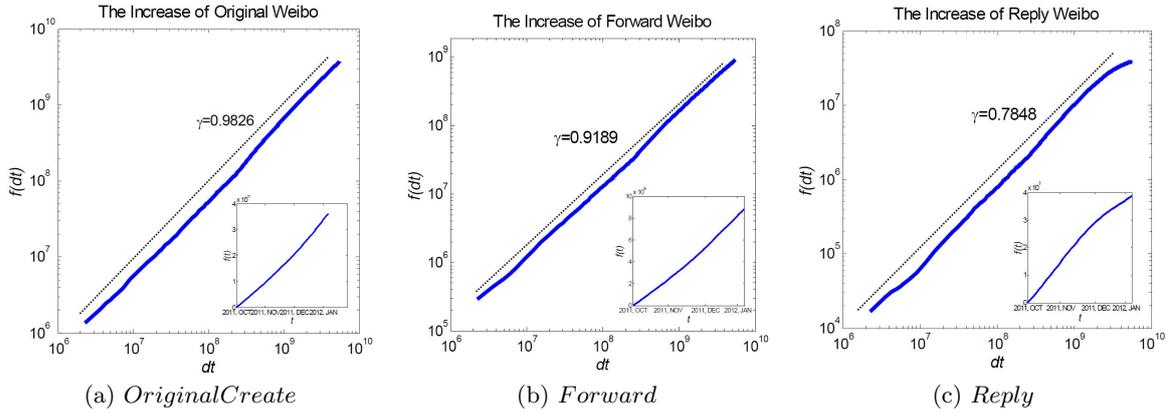

(a) *OriginalCreate*  (b) *Forward*  (c) *Reply*

**Figure 1: Macro Increase of "Original", "Forward", "Reply" along with intrinsic time**

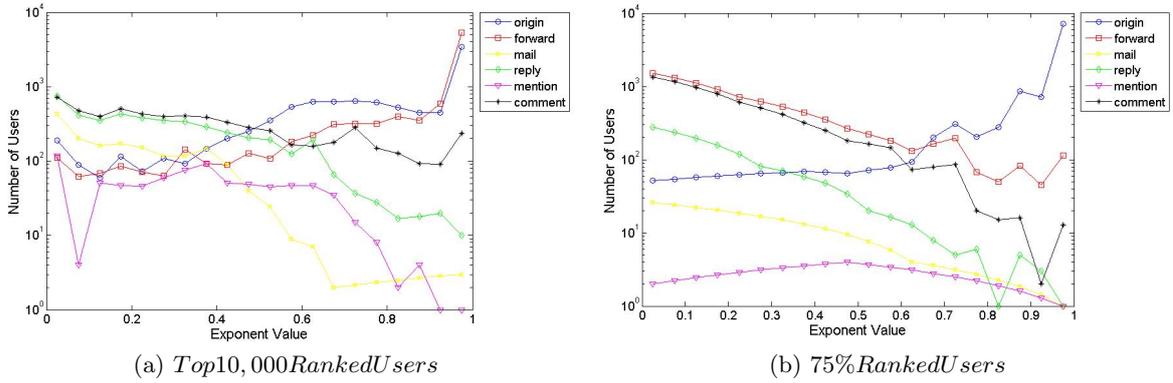

(a) $Top 10,000 Ranked Users$  (b) $75\% Ranked Users$

**Figure 2: Exponent Distribution of Users' Behaviors**

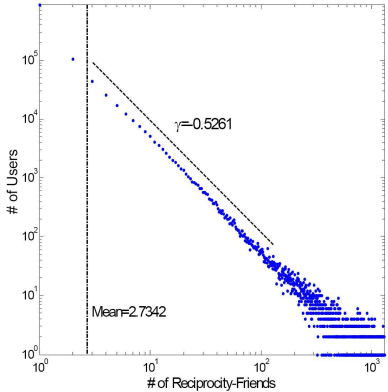

**Figure 3: The Distribution of Reciprocity Relationship**

### 4.2 Information Diffusion

#### 4.2.1 Distribution of "Forwarding"

We draw "Forward" distributions in Figure 4, where the $X$ axis represents the number of "Forwards", while the $Y$ axis represents the number of weibos that have a common number of "Forwards". As seen in Figure 4, the curve satisfies power-law distributions with the exponent -1.7415, which is bigger than that of Del.ici.ous [5] at -3.5, Youtube and Flickr [15], which are -3.5 and -8.2 respectively(they count how many tags are generated for one certain resource). The mean value is 10.0304, which means that for each weibo in Tencent, the average number of forwards is around 10. This value is also bigger than Del.icio.us, Youtube and Flickr. As for Twitter, according to our Twitter data set, the average value of retweets is about 2.3609, which is far more smaller than Tencent; besides, only 6% tweets in Twitter are retweeted, while 16.7% weibos in Tencent are forwarded, this result show that information in Tencent can reach up to a wider range than that in Twitter.

Contrasting to the high average forwarding value, 97% weibos are forwarded less than 10 times, which means Tencent users have greater enthusiasm for popular news, and thus a small number of weibos have been forwarded by amazing number of times. This phenomenon is similar with Twit-ter.

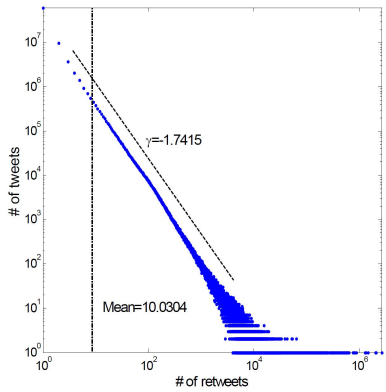

**Figure 4: The Distribution of Forward Behaviors**

#### 4.2.2 Depth of "Forwarding"

We draw the depth distribution of "Forward" propagation in Figure 5. As seen in this Figure, 96% Tencent weibos have never been forwarded, 95% forwarded weibos have been forwarded less than 10 times, while for the rest of weibos, the depth distribution satisfies power-law distribution with a decrease exponent as -2.8599, which is smaller than Twitter, which is -1.5114 [14]. The average "Forwards" depth is 1.2898, which is similar with that of Twitter. The deepest length is 69, which is far bigger than that of Twitter, which is less than 20. Only 75 weibos reach that level. There are two possible reasons for that, first, The follow network in Tencent may be more complex than that of Twitter; both micro-blogs have similar number of users at the end of 2011, while the average "Following" number of Tencent is around 64, which is higher than that of Twitter, which is less than 50; the high activities of Tencent users to follow others will cause higher complexity. second, it is more general in Tencent that there exists many online business groups to heat up certain events with sensitive factors (most of which are related with public impartial, moral and etc) to attract public's eyes, so that they can make profit from it; one main method is to keep forwarding the related information in a very short time period, which would also increase the depth of Forwards.

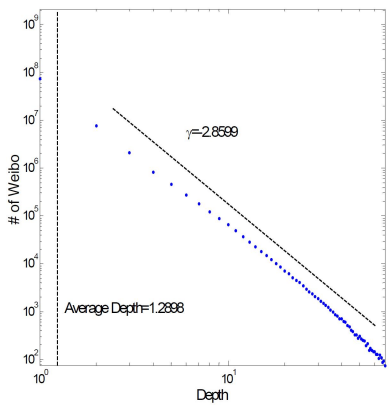

**Figure 5: The Relationship between "Follows" and "Forwards"**

#### 4.2.3 Time of "Forwarding"

As seen in Figure 6, about 35 % Forward actions happened in one hour, 45 % happened in one day, 14% happened in one week and 4.3% happened after one month. The distribution satisfies power-law distributions, where the exponent is -1.9058 and the average forwarding time is 95,875 seconds, which is about 26 hours.

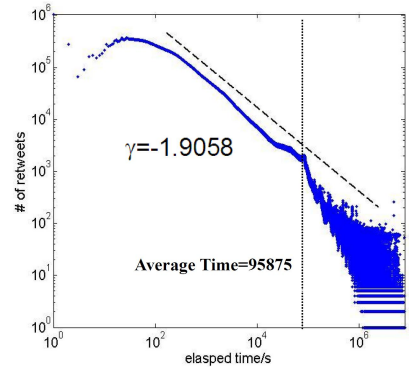

**Figure 6: The Relationship between "Forward" Propagation and "Time"**

Compared with Twitter, the speed of information spread in Tencent is a little slower in one hour (10% slower), while faster in one day (8% faster) and one week (12% faster). The reason for this is that the networks generated by Tencent users are more complex than that of Twitters, which make the spreading speed less efficient than that of Twitter; in another aspect, Tencent users have a higher rate of propagating information than Twitter users, thus after a time period, information in Tencent often spreads to a wider range than Twitter.

### 4.3 Topical Analysis

#### 4.3.1 Ranking Topics and Tencent Users

In Tencent Weibo, users can also create topics, which is similar to *hastag* in Twitter. Those created topics help us better understand the daily interests of Tencent Users. We first collect all created topics for each day (we only consider topics with participants), where we obtain around 50,000 of the most popular topics. We then select the top 20 ranked weibos for each topic as its content, and run the data by applying parallel LDA [2] model to make clustering on all topics(we assign the number of Topics as 50, and exhibit the top 10 ranked topics). The topic distribution of the top 10 days (From 7 Oct 2011 to 17 Oct 2011), second 10 days and third 10 days can be observed in Figure 7:

As seen in Figure 7, different with Twitter, the weight of entertainment (such as Online Game, QQ Space, Picture and Video Share) is highest in Tencent Weibo; besides, a certain amount of users are constantly interested in Sports, Cars and Economics, and keep creating related weibos at all times. One interesting phenomenon is that users are more interested in suddenly happening events, most of which are related to social affairs and global events. For example, in Time Period 2, a transportation accident and a homicide event in the Golden Triangle brought up a huge wave in

---
[2] http://code.google.com/p/plda/

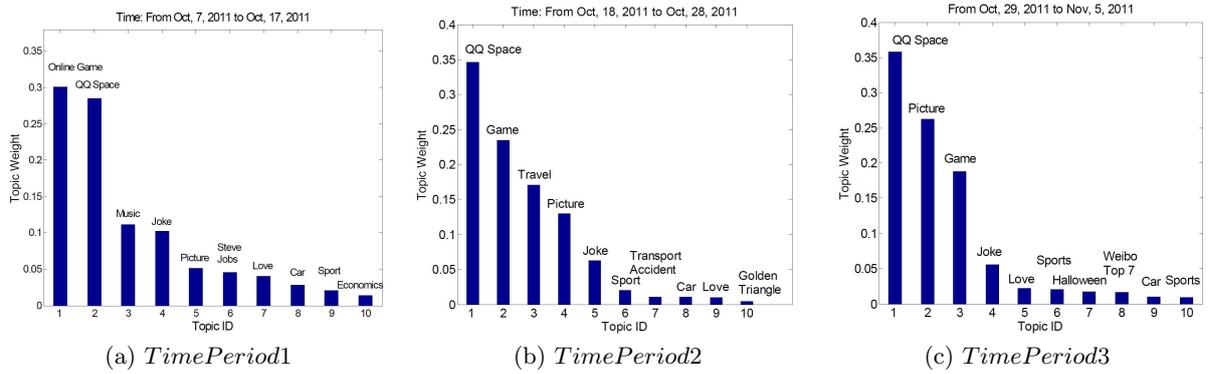

Figure 7: Topic Distributions for Tencent Weibo From Different Time Periods

Tencent Weibo, and the rank of related key words got into top 30 only in 10 days. This means that a huge amount of users in Tencent Weibo are more interested in specific topics, and they can share and spread certain information quickly to rapidly propagate it to a wide range.

We also rank all Tencent users by applying a parallel PeopleRank algorithm, which is an improvement version based on parallel PageRank [12]. We compare the results with Twitter as seen in Table 2.

In Table 2, the top 20 ranked Tencent users are mainly in entertainment, sports, and are famous hosts, actors, sports and the Tencent Official Agency itself; a famous writer and economist are also included in the top 20 ranked Tencent users; while in Twitter [14], the role of most of the top ranked users are similar to Tencent, while the first ranked is *BarackObama*. The main difference is that in Tencent Weibo, the number of top ranked official Agencies (nine Agencies) are bigger than that of Twitter (three Agencies), which means that traditional media still plays an important role in broadcasting information and providing services in Tencent Weibo. In another aspect, different with Twitter, user accounts with content related with jokes, literatures and fashions take a high proportion(about nine out of top twenty ranked accounts), this phenomenon is mainly based on different culture back ground, while in China, people often like to use poetry, joke to express their sentiment or dissatisfies; for young people in China, they have high enthusiasm to pursue fashions.

### 4.3.2 Topical Trending Analysis

First, we investigate what kind of information could be "Popular" and their popular patterns in Tencent Weibo. We find four different popular patterns from all detected topics, where sampled topics for each pattern are drawn in Figure 8. Each time period contains 10 days and the influence weight is the normalization of their weibos.

Based on Crane and Sornette's theory [7], popular patterns can be divided into four categories, which can be responses to the four sampled figures. The first is mainly determined by *exogenous*, which can be seen as a sudden outbreak, rapidly spread throughout the whole Tencent Weibo in a very short time, and then decreases to a low level for a long time period; "School Bus" is a serious accident, which caused public's attentions towards Government management for quite a long time. The second is determined by both *exogenous* and *endogenous*, where *exogenous* is seen to have a big influence on *endogenous*. For example, the event of *Gaddafi* lasted for almost one year, and users still kept a certain active degree to talk about this topic, when the "Death of Gaddafi" ourbreak, the popularity reached an amazing height and continued for several days, after that, the popularity of this topic decreased rapidly and kept in a lower level. The essence of this pattern is that if users' expectations are satisfied by external factors, then their concerns towards certain event will become lower. The third pattern is the circle of the second one, especially in stock market, users' expectation will be satisfied again and again by external factors and the *Fluctuation* will be generated. The fourth pattern exhibits a process of life circle, this pattern is especially general for online games, products and etc, which can attract and keep a huge amount of users to receive their services for quite a time period, and after that, the influence degree will decrease slowly. As can be seen in Sub-Figure (d), the game *SkyScraper* is very popular during October and December, the game reached its top value when it came to the end of November, after that, Tencent users exhibited less and less interests in the game.

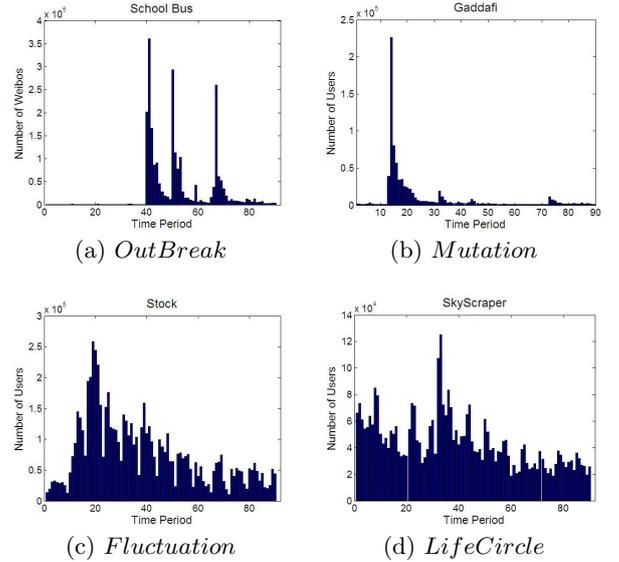

Figure 8: Popular Patterns in Tencent Weibo

Table 2: Top 20 ranked users by PeopleRank from both Tencent and Twitter

| Rank | Tencent Ranking | | | Twitter Ranking | | |
|---|---|---|---|---|---|---|
| | ID | Name | Remark | ID | Name | Remark |
| 1 | 1323005700 | High Quality Joke | Services | aplusk | ashton | actor |
| 2 | 30818627 | We like Jokes | Services | obama | Obama | president |
| 3 | 88886666 | Super QQ | Official Weibo | CNNBrk | CNN Breaking News | news |
| 4 | 970144221 | DNF | Official Online Game | TheEllenShow | Ellen DeGeneres | Show host |
| 5 | 2360330217 | Mood Helper | Official Weibo | britneyspears | Britney Spears | musician |
| 6 | 19990210 | QQ Product Team | Official Weibo | Oprah | Oprah Winfrey | show host |
| 7 | 622004906 | Na Xie | Host | THE REAL SHAQ | THE REAL SHAQ | Sports Star |
| 8 | 611986579 | Jiong He | Host | Johncmayer | John Mayer | Musician |
| 9 | 1379986183 | Beauty Story | Official Weibo | twitter | Twitter | Twitter Weibo |
| 10 | 2367520831 | Libo Zhou | Show Host | RyanSeacrest | Ryan Seacrest | Show Host |
| 11 | 302536308 | Text that touch your soul | Lancearmstrong | Lance Armstrong | Sports Star |
| 12 | 3756217 | Hot Jokes | Services | JimmyFallon | Jimmy Fallon | Actor |
| 13 | 2581425470 | QQ Net | Services | Iamdiddy | Iamdiddy | Musician |
| 14 | 611985987 | NBA | Sports | muskutcher | Demi Moore | Actress |
| 15 | 305974257 | Weian Qiao | Writer | PerezHilton | Perez Hilton | Power Blogger |
| 16 | 345005673 | Beauty Sentences | Services | nytimes | The New York Times | News |
| 17 | 622006051 | Zhiqiang Ren | Finance | mileycyrus | Mile Ycyrus | Actress and Musician |
| 18 | 343551325 | Joke Base | Services | stephenfry | Stephen Fry | Actor |
| 19 | 772937290 | Fashion and Constellation | Services | The Onion | The Onion | News |
| 20 | 99912345 | Tencent News | News | KimKardashian | Kim Kardashian | model |

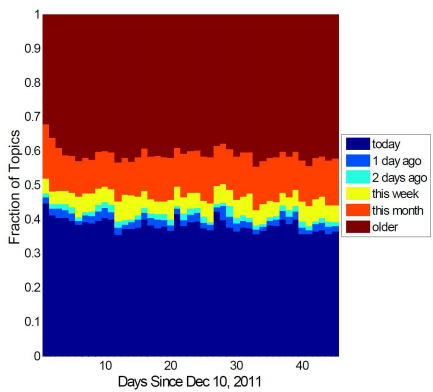

**Figure 9: The Age Trending of Topics in Tencent Weibo**

In another aspect, we observe the age of trending topics (The Freshness of Topics) from Tencent and compare it with Twitter and Google [14]. The results of Tencent seen in Figure 9 exhibits the proportion of old and fresh information at any time slice.

In Tencent, the proportion of older topics (more than one month) is around 40 %; in Google, the proportion of older keywords is less than 10%, while in Twitter, the proportion of older tweets is less than 20% [14]. The phenomenon shows that many older topics in Tencent Weibo are consistently popular and discussed among Tencent users, where those topics provide steady platforms for users to know and communicate with each other; on the other hand, the proportion of new topics in each day is also around 40%. This phenomenon thus illustrate different behavior patterns between Tencent and Twitter, in Tencent, many celebrities like to create hastags such as real estate, situation of middle east as a micro forum, and a lot of other users would like to join in it for a long period of time.

## 5. MICRO LEVEL ANALYSIS

In this section, we focus on the relationship between users' actions and their social influence from micro level, and compare the result with Twitter. In our researches, a user's influence is mainly measured from two aspects: "Followers" and "Followees", "Forward" actions. To better evaluate users' influence from these two aspects, a batch of experiments are designed for each question, where the results summary is presented in the following sections.

### 5.1 Analyze Influence from "Follow" Actions

In this section, we identify three subsections of the "Follow" actions: the first subsection is to introduce the statistical features of "Followees" and "Followers", which could also be used to describe the influence distribution of Tencent users; the second subsection is to detect how the number of "Followees" will influence the number of "Followers"; the third subsection is to detect how other actions will influence the number of "Followers".

#### 5.1.1 Distribution of "Followees" and "Followers"

We draw the distribution of "Followees" and "Followers" in Figure 10, where both satisfy power-law distributions with steep power-law tails(the exponent of "Followees" is -3.0096 and the exponent of "Followers" is -3.6849, which are smaller than Twitter (all are around -2.276) [14] and similar to Delicious (around -3.5) [5]). The average "Followees" is 64, which is bigger than that of Twitter (around 42 according to our collected Twitter data) and the average "Followers" is 155, which is also bigger than that of Twitter (around 136 according to our collected Twitter data). The summarized data illustrates that Tencent users have higher activities to follow others, while they seem to follow a wider range than Twitter. This phenomenon also implies that Tencent may have a more complex network than Twitter.

As seen in Figure 10, the curve for "Followees" has two significant glitches, where one is around 15 and the other is around 40. The reason for this is that Tencent Weibo is developed based on Tencent QQ (the biggest online chatting services in China), so when users entered Tencent Weibo, they also brought some of their original relationships into Tencent Weibo, where the average amount is around 40; otherwise, if they did not bring any relationships, Tencent Weibo will recommend them 10 to 20 users according to their history records from Tencent QQ. Besides, more than 30 Tencent users obtained bigger than $10^5$ followers during the three months studies, similar to Twitter, those users are mainly celebrities (entertainment stars, hosts and economists), government agencies, famous companies and big media or entertainment agencies (such as Tencent News, on line game, joke center, beauty and fashion). These high ranked celebri-ties often have a high influence, and their messages will often be forwarded by tens of thousands of their followers, some of those celebrities, government agencies and companies would also like to follow some ordinary users, those Reciprocity behaviors often significantly improve the influence of ordinary users.

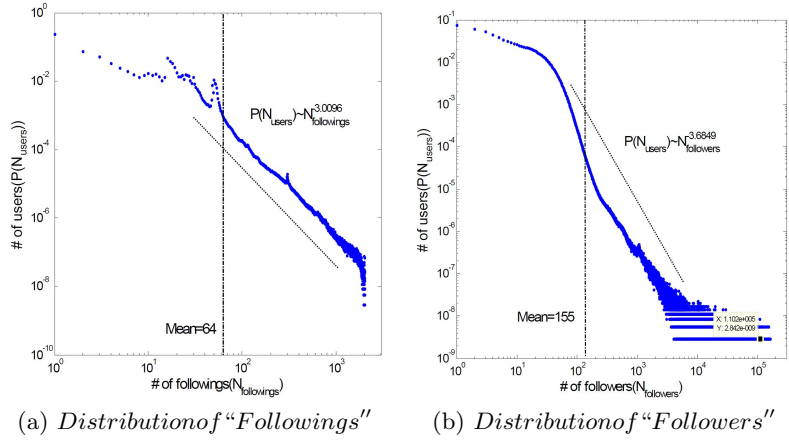

(a) $Distribution of\ "Followings"$  (b) $Distribution of\ "Followers"$

**Figure 10: Distributions of "Followings" and "Followers"**

### 5.1.2 Relations of "Followees" and "Followers"

In order to find the correlation between "Followees" and "Followers" of Tencent users, we plot the number of "Followees" in $X$ axis, the user has followed against the number of "Followers" in $Y$ axis, the user has obtained. As seen in Figure 11. The experiment data covers the whole range of users, which is about 320 millions.

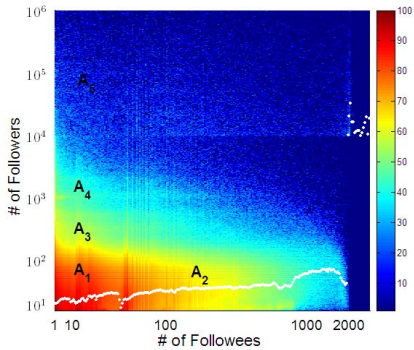

**Figure 11: Distribution of Followees-Followers**

As seen in Figure 11, the color points (total $800 \times 800$ color points) in the heat-map mean identifies the number "$N$" of users who have followed "$X$"(the number on X axes) users and have "$Y$" (the number on Y axes) "followers", we assign the points with maximum number of users as 100%, which is about 6,067,947. In order to clearly observe the distribution of users' "Followees" and "Followers", we make double log for all $N$; then all the users can be classified into five significant groups ($A_1,A_2,A_3,A_4,A_5$). Around 97% of users are in $A_1$ area, whose "followees" are from 5-400 and "followers" are from 10-400; around 2.6% users are in $A_2,A_3$ and $A_4$ areas; less than 0.4% users are in $A_5$ area. The vertical direction represents users' influences, while the horizontal direction represents users' activities, a bigger value of $Y$ therefore means that the user has a higher influence in obtaining more followers, while a bigger value of $X$ means that the user has a higher activity in following others. The number of users from $A_1$ to $A_5$ follows an exponent decrease with exponent around -6.3. The size of each area represents the diversity of users' followee-follower relationships, where for area $A_5$, though the total number is less than 0.4%, users' "Followee-Follower" relations covers a wide range. The curve in Figure 11 shows the increase of median value of each number of "Followees", different with Twitter, the trend from 0 to 1,500 exhibits a linear monotonous increase with the slope around 0.45, while when the number of a user's followees exceeds 1,500, he/she will have a higher probability to obtain more than 10,0000 followers, one main reason is that 75% of those users are official agencies, which pay more attention to the feedback of users' experiences for their services.

### 5.1.3 Relations of "Followers" and Other Actions

In this section, we investigate the relationships between the number of users' "Followers" and their total number of actions. As seen in Figure 12, the color nodes with maximum number of users are assigned as 100%, which is about 2,750,930. Five significant areas could also be observed, where $A_1$ takes up 95% of Tencent users, the number of users from $A_1$ to $A_5$ follows an exponent decrease with the value of exponent around -6.8. Different than Figure 11, the same number of actions often cover a wider range of "Followers", especially for the area $A_3$ and $A_4$. This means the diversity of users' actions has an influence on their number of followers; for example, assume two users have the same number of actions, where one may have far more followers than the other, because his contents are more interesting, or he has connected with more influential users than the other. For the majority of users, high quality content may help contribute 100 to 200 more followers than others.

In both Figure 11 and 12, The curves of median value keep a linear increase from $A_1$ to $A_4$, while they scatter in area $A_5$. This phenomenon shows that users' actions could help increase the number of followers to a certain extent, where 99% users are located in area $A_1$ to $A_4$. Very few users could exceed that extent, while for those users in $A_5$, it significantly shows that users' actions have no influence on the number of their followers. This phenomenon can be widely observed in Tencent Weibo, for example, many Tencent users say that they can not increase their followers any more by trying different methods. For a famous person, however, some of their casual messages could cause tens of thousands of forward,comment in less than an hour. This phenomenon can be explained as for a potential friends circle for each users, the scale of the circle is mainly based on the inner attributes of the user; these inner attributes will determine how many other users will be interested in current users. For example, a scientist can only attract those who are interested in his/her research domain, while a movie star has more common attributes that may attract more followers.

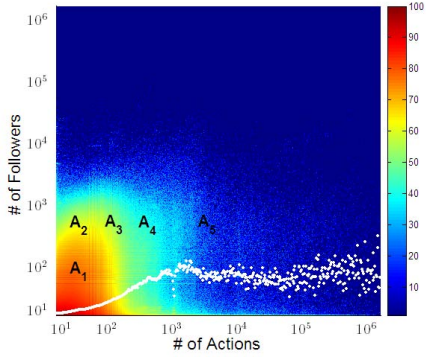

**Figure 12: The Distribution of Followers-Actions Relationship**

Similar results can also be observed in Twitter [14], but they exhibit different slope values. For example, the slope between "Followees" and "Followers" is around 1.24 in Twitter, while it is up to 3.21 in Tencent; the slope between "Followers" and "Actions" is around 0.45 in Twitter, while it is about 0.16 in Tencent. The phenomenon shows that in Twitter, users' number of actions will have a more significant effectiveness on the number of "Followers" than that of Tencent.

## 5.2 Analyze Influence from "Forward" Actions

In this section, we analyze users' "Forward" actions from two aspects, the first is to analyze relations between "Forward" and "Follow"; the second is to analyze relations between "Forward" and other actions.

### 5.2.1 Relations of "Forward" and "Follow"

Previous studies [14] have analyzed the distribution of "Retweets" behaviors of users' followees and followers. Based on their researches, we build up a similar experiment to analyze users' "Forwards" behaviors in Tencent Weibo. First, assume user $i$ has $k_1$ followers and $k_2$ followees, where to clearly illustrate the problem, we define $k_1 = k_2 = k$, after which a formula based on [14] is introduced as follows:

$$Y(i) = \sum_{j=1}^{k} \{\frac{\theta_{ij}}{\sum_{o=1}^{k} \theta_{io}}\}^2 \quad (2)$$

$\theta_{ij}$ means the number of times $u_j$ forwards $u_i$'s weibo ($u_j$ is a follower of $u_i$), or the number of times $u_i$ forwards $u_j$'s weibo($u_j$ is a followee of $u_i$). The formula could measure the distribution of $u_i$'s followers' "Forwards" behaviors towards him and $u_i$'s "Forwards" behaviors towards his followees. The results are seen in Figure 13:

In Figure 13, "In Degree" means all the "Forwards" actions of user $u_i$'s followers, while "Out Degree" means all the "Forwards" actions of user $u_i$ towards all his followees. The increase exponent of dash pitched line for "In Degree" is 0.9584(bigger than Twitter, which is 0.801) and 0.8828 for "Out Degree"(smaller than Twitter, which is 0.892). According to Kwak's analysis, the more closer to the dash line, users' most of forwards will have a higher probability occurs within a subset of their followers; the more closer to $X$ axis, users will obtain a more even distribution of their "Forwards" actions. The exponents show that compared with Twitter, some celebrities or agencies have received more attentions and have more influence in Tencent; while for users, their focus on a subset of their followees are not as strong as that in Twitter. The median curve in two sub-figures also show that Tencent users would like to forward others' weibo in a more wider range.

### 5.2.2 Relations of "Forward" and Other Actions

In this section, we would like to investigate whether the influential relationship of two users can be measured. We take "Forward" as the main index to evaluate the influence between two users $X_1$ and $X_2$, if $X_1$ has a high probability to forward $X_2$'s weibos, we call that $X_2$ has a high influence on $X_1$. We mainly consider three features: The number of replies, comments and mails between $X_1$ and $X_2$. We take the three features as X-axis, the forward probability between random users pairs as Y-axis, and draw Figure 14.

In Figure 14, the linear parts of each sub-figure take up 99% users(their "Reply" or "Mail" actions are less than 20 and 45 times respectively), which mean that we could directly summarize the relation patterns from the linear parts. As can be seen, "Replies" and "Mails" have positive correlations with "Forward" probabilities, but "Comments" do not have significant relations. The reason is that users' comments have randomness, because they often comment on a certain weibo and do not care who provides it.

## 6. APPLICATIONS

The analysis above have its practical value in reality, especially for behaviors prediction. In this section, We take "Forward" anlysis in the previous section as an example, and propose a Conditional Random Field(CRF) based prediction model to verify the application values of our analysis. we select the number of "Replies", "Mails" and "Comments" as three features, we also add other features such as keywords, topics of weibos as aids to further improve the accuracy. we use exponent increase function to simulate the relationship between the selected features and "Forward" probabilities, which could be seen as below:

$$P(y|n_{type}(X_1, X_2)) = \frac{e^{n_{type}}}{1 + e^{n_{type}}} \quad (3)$$

$y$ represents whether user $X_1$ forwards $X_2$, $y = 1$ means to forward and $y = 0$ means to not forward; $type$ means different actions, which include "Reply", "Mail" and "Comment", $n_type$ means the number of times $X_1$ "Replies" or "Mails" or

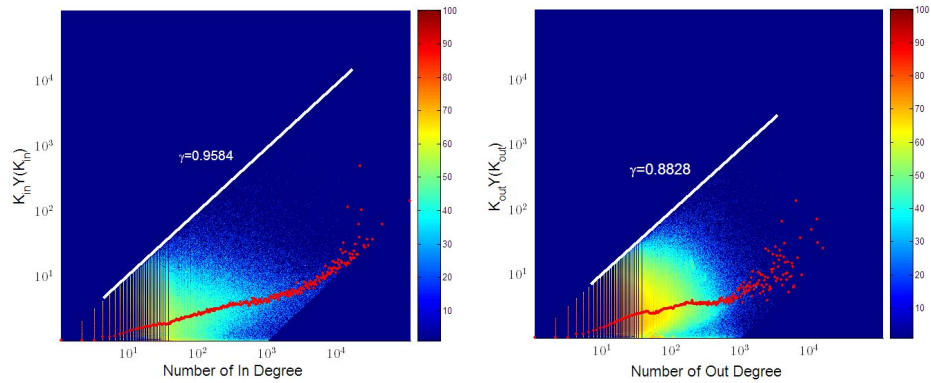

Figure 13: The Relationship between "Follows" and "Forwards"

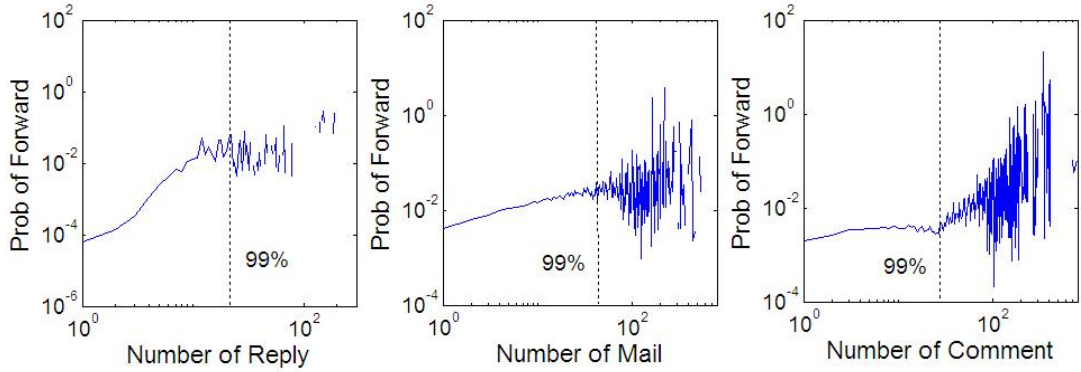

Figure 14: Relations between "Forward" Probability and Other Actions

Table 3: Contribution of different indexes for "Forwarding" prediction

| Item | All | No Reply | No Mail | No Comment |
|---|---|---|---|---|
| Accuracy | 0.6902 | 0.6746 | 0.6702 | **0.6903** |
| Recall | 0.9605 | 0.8958 | 0.8884 | **0.9605** |
| F1-Score | 0.8088 | 0.7883 | 0.7875 | **0.8089** |

"Comments" on $X_2$. $P(y|n_type(X_1, X_2))$ is the probability distributions of $X_1$ to forward $X_2$ based on the features of $n_type(X_1, X_2)$. Experiments data are collected from all actions of 2,000 high level active users, and the results can be seen in Table 3

Experiment shows that "Reply" and "Mail" provide positive contributions towards the performance, while "Comment" provides negative contributions towards the performance. The result is consistent with our analysis; it also illustrates that statistical analysis for Tencent weibo, could help to detect the inner relations of users' influence and is a good guiding significance for system design and optimization.

## 7. CONCLUSION

The paper proposes a systematic study on Tencent Weibo, one of the largest Micro-Bloggs in China, from both macro and micro level, and compare the results with Twitter. From Macro level, Tencent Weibo is more like a personalized media center, the network of Tencent is more complex than Twitter, and the users are more active than Twitter; Besides, the topics, which are talked in Tencent are very different with that of Twitter, Tencent users like topics related with entertainment, joke and fashion, they also like to keep communicating on a certain number of old topics, which were marked by other users. The unique features of Tencent users' behaviors are mainly determined by the culture background in China. From Micro level, we mainly investigate users' social influence from two aspects: "Follow" and "Forward", we first analyze the statistical features of "Follow" and "Forward", then we research on the relations between "Follow", "Forward" and other users' actions, we find that "Follow" and "Forward", "Follow" and users' actions have positive relations under a certain extent; some of users actions(Reply, Mail) have significant positive relations with "Forward" probabilities, while "Comment" do not have that relation. At last, a predict model is proposed to verify the application value of our analysis. Experiment results show that the analysis could help to better understand micro-bloggs and provide good guiding significances for system design and optimizations.